\documentclass[aps,amsmath,amssymb,preprintnumbers,a4paper,prd,nofootinbib,onecolumn]{revtex4-1}
 \pdfoutput=1
 
\usepackage{lipsum}
\usepackage{bm}
\usepackage{mathrsfs}  
\usepackage{slashed}
\usepackage{subfigure}
\usepackage{feynmp}
\usepackage{graphicx}
\usepackage{amsfonts}
\usepackage{color}
\usepackage{cancel}
\usepackage{amsmath}
\usepackage{makecell}

\def\Tr{\mbox{Tr}\,}

\def\hbar{\hspace{0pt}\raisebox{1pt}{$-$} \hspace{-7pt} h}

\def\5{\overline 5}

\definecolor{JJ}{RGB}{0,144,255}

\newcommand{\be}{\begin{equation}}
\newcommand{\ee}{\end{equation}}
\newcommand{\bea}{\begin{eqnarray}}
\newcommand{\eea}{\end{eqnarray}}

\newcommand{\ba}{\begin{eqnarray}}
\newcommand{\ea}{\end{eqnarray}}

\definecolor{rossoCP3}{cmyk}{0,.88,.77,.40}
\definecolor{graa}{rgb}{0.8,0.8,0.8}
\definecolor{blaa}{rgb}{0.2,0.2,0.6}

\begin{document}
\title{\textcolor{rossoCP3}{Collider Tests of (Composite) Diphoton Resonances}}

\author{Emiliano Molinaro}

\author{Francesco Sannino}

\author{Natascia Vignaroli}
 \affiliation{CP$^{3}$-Origins and the Danish IAS, University of Southern Denmark, Campusvej 55, DK-5230 Odense M, Denmark}

\begin{abstract} 
We analyze the Large Hadron Collider  sensitivity to new pseudoscalar resonances decaying into diphoton with masses up to scales of few TeVs. We focus on minimal  scenarios where the production mechanisms involve either   photon or top-mediated gluon fusion,  partially motivated by the tantalizing excess around 750 GeV reported by ATLAS and CMS. The two scenarios lead respectively to a narrow and a wide resonance.  
We first provide a model-independent analysis via effective operators and then introduce minimal models of composite dynamics where the diphoton channel is characterized  by their topological sector. The relevant state here is the pseudoscalar associated with the axial anomaly of the new composite dynamics.  If the Standard Model top mass is generated via  four-fermion operators  the coupling of this state to the top remarkably explains the wide-width resonance reported by ATLAS.   
Beyond the excess, our analysis paves the way to test dynamical electroweak symmetry breaking via topological sectors.

\preprint{CP3-Origins-2016-008 DNRF90, ~ DIAS-2016-8 ~~~~}
 
\end{abstract}

\maketitle

\section{The relevance of the Diphoton channel}

The diphoton channel has proved extremely successful in discovering new (pseudo)scalar particles such as the Higgs boson
\cite{Aad:2012tfa,Chatrchyan:2012xdj}. Earlier $\pi_0 $ and $\eta^{\prime}$ decays into $\gamma \gamma$ provided instrumental to demonstrate the composite nature of the QCD hadrons. Furthermore, a tantalizing excess around 750 GeV has been reported by the LHC experimental collaborations \cite{ATLAS,CMS} in the current run at $\sqrt{s}= 13$ TeV. More specifically, with $3.2$ fb$^{-1}$, ATLAS observes an excess in the number of events, respectively of about $7.8$ and $4.3$, for the diphoton invariant mass bins at 730 and 770 GeV. This corresponds to a local significance of 3.9$\sigma$ at 750 GeV, under the assumption of a large width of about 45 GeV. With 2.6 fb$^{-1}$ CMS measures an excess around 760 GeV, corresponding to a local significance of $2.6\sigma$.

It is therefore timely to investigate the LHC reach and constraints from the diphoton channel. 
Because of the intriguing excess around 750 GeV, we first specialize our analysis around this energy range. Minimal models that can explain the excess entail gluon fusion production through new colored states \cite{Molinaro:2015cwg, Pilaftsis:2015ycr, Franceschini:2015kwy, DiChiara:2015vdm, Antipin:2015kgh, Knapen:2015dap, Dev:2015vjd} or photon fusion production \cite{Fichet:2015vvy,Csaki:2015vek,Csaki:2016raa,Harland-Lang:2016qjy} that  typically lead to a narrow resonance (see also \cite{Godunov:2016kqn,Djouadi:2016eyy}). If a wide-width scenario, currently favored by ATLAS \cite{ATLAS}, is confirmed, it can be achieved via a direct coupling to the top quark \cite{Bellazzini:2015nxw, Franzosi:2016wtl}. This induces the production of the pseudoscalar resonance via top-mediated gluon fusion.  
 Alternative ways to obtain a wide width are via exotic decay topologies \cite{Bernon:2015abk, Cho:2015nxy, Huang:2015evq, Chala:2015cev, Aparicio:2016iwr, An:2015cgp} or invoking a coupling to an invisible sector \cite{Mambrini:2015wyu, Bi:2015uqd, D'Eramo:2016mgv, Redi:2016kip}. 
 
 Here we therefore consider two production mechanisms, the photon and top-mediated gluon fusion. We will first rely on an effective field theory approach and then consider minimal models of composite dynamics. 
We discuss the current constraints and excesses coming from the LHC run at $\sqrt{s} =8$ TeV (LHC-8) as well as the run at  $\sqrt{s} =13$ TeV (LHC-13).  We observe that when the resonance is photo-produced one can constrain it up to high mass values of the order of $5$~TeV with around $100~{\rm fb}^{-1}$. 
If the diphoton resonance is produced via top-mediated gluon fusion the reach is up to $2$~TeV with  $100~{\rm fb}^{-1}$, due to the quick drop of the production cross section with its mass.  

Minimal models of composite dynamics all predict a pseudoscalar state with specific couplings to the electroweak (EW) gauge bosons which arise from the topological sector of the underlying theory \cite{DiVecchia:1980xq,Molinaro:2015cwg}. This state is the analogue of the $\eta^\prime$ of QCD. This makes the models ideal case-studies for the diphoton channel. The composite pseudoscalar resonance also offers a natural explanation for the observed excess at 750 GeV, as shown in \cite{Molinaro:2015cwg}. Other composite realizations have been explored in \cite{Harigaya:2016pnu,Kim:2015xyn,Barrie:2016ntq,Franzosi:2016wtl,Liao:2015tow,No:2015bsn,Harigaya:2015ezk,Cline:2015msi, Belyaev:2015hgo}. 

We first review the effective Lagrangian for minimal models of composite dynamics augmented by the gauged version of the Wess-Zumino-Witten term \cite{Witten:1983tx,Wess:1971yu,Kaymakcalan:1984bz,Kaymakcalan:1983qq,Schechter:1986vs,Duan:2000dy}.  We then move to its phenomenology in the two limits of photon and gluon fusion production of the  $\eta^\prime$-like state.  We will see that if the Standard Model (SM) top mass is generated via four-fermion operators, the coupling of this state to the top naturally explains the wide-width resonance reported by ATLAS.

Our results demonstrate that topological sectors stemming directly from the underlying dynamics give rise to novel signatures in the diphoton and EW channels that open a new avenue to test natural theories of  EW  symmetry breaking at present and near future collider experiments. In that respect, our analysis complements phenomenological studies of composite dynamics that, so far, have been mostly focused on spin-one resonances \cite{Aad:2015owa,CMS:2015gla}.

The paper is structured as follows:  in section~\ref{sec:xsec} we set up the analysis  via an effective operator framework and study the photon and the top-mediated gluon fusion production mechanisms. We also compare with current collider limits and discuss the observed excesses at 750 GeV. We then determine the LHC-13  reach for higher masses.  In section~\ref{sec:model} we introduce the minimal models of composite dynamics and their effective Lagrangian including the topological terms. For the two envisioned production mechanisms we analyze the LHC-13  reach and constraints stemming from these terms. We finally offer our conclusions in section~\ref{summary}.  Further details related to the topological terms can be found in the appendix.

\section{LHC-13 Reach on Diphoton Resonances}\label{sec:xsec}

We  derive the LHC-13 reach on diphoton resonances via an effective approach   in the two hypotheses of dominant photon and gluon fusion production mechanisms. Clearly the final result depends sensitively on how the new physics couples to SM degrees of freedom. We encode the new physics in the effective theory below: 
\begin{align}\label{eq:L-eff}
\begin{split}
\mathcal{L}_{\rm eff}=& -i y_t \frac{m_t}{v} a\, \bar{t} \gamma_5 t - \frac{c_{GG}}{8v}  a\, \Tr\left[ G^{\mu\nu}\tilde{G}_{\mu\nu}\right] \\
& - \frac{c_{AA}}{8v}  a\, A^{\mu\nu}\tilde{A}_{\mu\nu} - \frac{c_{AZ}}{4v}  a\, A^{\mu\nu}\tilde{Z}_{\mu\nu}- \frac{c_{WW}}{4v}  a\, W^{+ \mu\nu}\tilde{W}^{-}_{\mu\nu} - \frac{c_{ZZ}}{8v}  a\, Z^{\mu\nu}\tilde{Z}_{\mu\nu}\,,
\end{split}
\end{align}
in which $a$ is a new pseudoscalar boson of mass $m_a$ in the TeV energy range,  $v=246$ GeV is the EW scale, and $\tilde{V}^{\mu\nu}=\epsilon^{\mu\nu\rho\sigma}V_{\rho\sigma}$. Note that we have reabsorbed the scale of new physics $\Lambda_{\rm NP}$ into the definition of the effective couplings 
\begin{equation}\label{eq:c-lambda}
 c_{VV} \approx \frac{1}{4 \pi}\frac{v}{\Lambda_{\rm NP}} \ ,
 \end{equation}
 whose specific value depends on the underlying theory.  We neglect in $\mathcal{L}_{\rm eff}$ the direct couplings of $a$ to the SM fermions except for the top quark $t$. Taking into account the tree-level unitarity bound from the vector bosons scattering amplitudes, i.e. $VV \to V^{'}V^{'}$, one finds that the effective theory is reliable up to energies of the order $\sqrt{s} \sim m_a \lesssim \sqrt{4\pi}\left(4\pi \Lambda_{\rm NP} \right)\lesssim$ tens of TeVs for $\Lambda_{\rm NP}\geq v$.
 Furthermore, we do not include in (\ref{eq:L-eff}) dimension-6 operators which become important at energies $\sqrt{s}\sim m_a \approx 4 \pi\Lambda_{\rm NP}$. These terms typically imply corrections to $c_{VV}$ in (\ref{eq:c-lambda}) of the order $(m_a/(4 \pi \Lambda_{\rm NP}))^2$, that we find to be negligible in most of the relevant parameter space for the $a$ phenomenology at the LHC. \\
As the $c_{VV}$  related operators are non-renormalizable, they arise via either non-perturbative dynamics or loop corrections. This implies that these coefficients contain both information about new physics and also the calculable SM contributions coming from the renormalizable interactions of $a$ with SM particles, in this case the top quark. Hence, $\mathcal{L}_{\rm eff}$ can be viewed as a conservative but sufficiently general effective description of a new pseudoscalar state. In this simple picture, we will not invoke the presence of new colored vector-like states that can also serve to produce this state \cite{Molinaro:2015cwg, Pilaftsis:2015ycr, Franceschini:2015kwy, Antipin:2015kgh, Knapen:2015dap, Dev:2015vjd}. This means that, in our case, the effective coupling to the gluons $c_{GG}$ is entirely given by the top loop and reads 
\begin{equation}
c_{GG}= y_t \frac{\alpha_S}{2 \pi}F\left(\frac{m_a^2}{4\,m_t^2}\right) \ ,\label{cGG}
\end{equation}
where $m_t$ is the top mass, $\alpha_S$ is the strong coupling constant and $F(x)=-\frac{1}{4x}\left(\ln\frac{\sqrt x+ \sqrt{x-1}}{\sqrt x- \sqrt{x-1}}- i\pi\right)^2$ for $x>1 $ \cite{Steinberger:1949wx}. As for the other $c_{VV}$, the top loop contribution is included but we will keep them free to accommodate the effects of new non-colored states. In our analysis we will consider two limits, one in which the coupling to the top is order $y_t \simeq 1$, still within the perturbative regime, and the one in which  $y_t =0$. In the first case we will have a top mediated gluon fusion production of the new state $a$, while in the second we will have the photon production.  In the  case $y_t \neq 0$ the presence of the top, as we shall demonstrate, naturally enhances the total width of $a$ up to $\Gamma_{\rm{tot}}(a)/m_a \approx 0.06$. 
  
From (\ref{eq:L-eff}) the effective partial decay rates read:
 \begin{eqnarray}
 	&&\Gamma(a \to g g) =  \frac{m_a^3}{8\,\pi}\, \frac{\left|c_{GG}\right|^2}{v^2} \label{gammaGG} \\
	&&\Gamma(a \to \gamma\gamma) \,= \, \frac{m_{a}^3}{64\,\pi}\,\frac{c_{AA}^2}{v^2}\,, \label{eq:rateGammaGamma}\\
	&&\Gamma(a \to \gamma Z) \,= \, \frac{m_{a}^3}{32\,\pi}\,\frac{c_{AZ}^2}{v^2}\left(1\,-\,\frac{m_Z^2}{m_a^2}\right)^3 \, , \\
	&&\Gamma(a \to Z Z) \,= \, \frac{m_{a}^3}{64\,\pi}\,\frac{c_{ZZ}^2}{v^2}\left(1\,-\,\frac{4\,m_Z^2}{m_a^2}\right)^{3/2}\,  ,\\
	&&\Gamma(a \to W^+W^-) \,= \, \frac{m_{a}^3}{32\,\pi}\,\frac{c_{WW}^2}{v^2}\left(1\,-\,\frac{4\,m_W^2}{m_a^2}\right)^{3/2}\, \label{eq:rateWW} ,\\
	&&\Gamma(a \to t\bar{t}) \,= \, y^2_t \, \frac{3\,m_a}{8 \pi}\frac{m_t^2}{v^2}\, \sqrt{1-\frac{4\,m_t^2}{m_a^2}}\label{gammattbar}
 \end{eqnarray}
 
\begin{figure*}[t!]
\begin{center}
\includegraphics[width=0.5\textwidth]{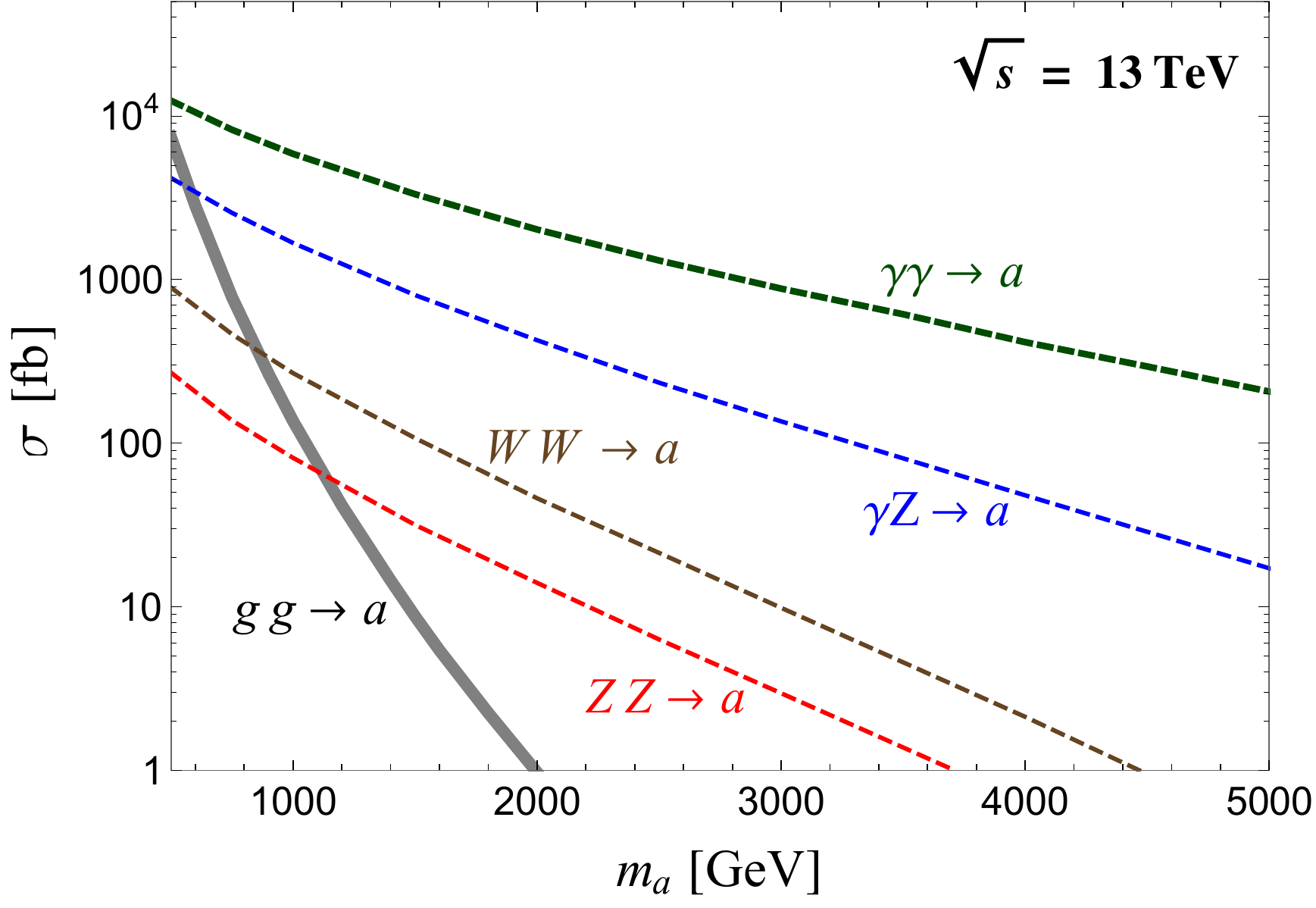} \,~~~~~~~~
\caption{The production cross section of the pseudoscalar resonance at  LHC-13 as function of its mass for the different modes: $\gamma\gamma$, $\gamma Z$, $WW$ and $ZZ$ fusion (with $c_{AA}=c_{AZ}=c_{WW}=c_{ZZ}=1$) and top-mediated gluon fusion (with $y_t=1$), see (\ref{eq:L-eff}). The gluon fusion cross section has been evaluated at NLO in QCD. The
photon fusion includes the dominant incoherent  as well as the subdominant semi-coherent and coherent 
contributions.}
\label{fig:xsec}
\end{center}
\end{figure*}

 As a logical step we start by plotting the production cross sections for various mechanisms relevant at LHC-13 as function of  $m_a$ stemming from our effective action. These are summarized in  Fig. \ref{fig:xsec}.  For illustration, 
 in the plot we assume $y_t=1$ for the gluon fusion production and $c_{VV}=1$ for the other processes.
 
The photon fusion production mechanism receives three contributions: the leading one (ranging from 60\% to 85\% for $m_a$ from 0.5 to 5 TeV) comes from incoherent/inelastic scattering, whereas two subdominant contributions arise from the semi-coherent and the coherent scattering processes \cite{Fichet:2013gsa,Fichet:2014uka}, where either one of or both the colliding protons remain intact. We determine the production cross section at leading-order (LO) with  MadGraph5\_aMC@NLO \cite{Alwall:2014hca}, by using the NNPDF2.3QED \cite{Ball:2013hta} set of parton distribution functions (PDFs). The improved Weizsaecker-Williams formula \cite{Budnev:1974de} is automatically employed by MadGraph5 to simulate low virtuality photon emission by proton beams. This allows to estimate the elastic and semi-elastic contributions \footnote{Our production cross section at $m_a=750$ GeV agrees with the results in \cite{Csaki:2016raa,Fichet:2015vvy}.}.  
The largest error on the cross section comes from the uncertainty on the photon PDF. In particular, for the NNPDF2.3QED set used in our analysis the quoted uncertainty is typically of the order of 50\% \cite{Ball:2013hta}, but it might  be even bigger in the large $x$ region, $x\gtrsim 0.1$ ($m_a \gtrsim 1.3$ TeV at LHC-13), due to the lack of experimental data. Our estimates are thus subject to $\mathcal{O}(1)$ corrections at large $m_a$ \cite{Ball:2013hta,Harland-Lang:2016qjy}~\footnote{For example, by using the other available MRST2004QED PDF set \cite{Martin:2004dh} we obtain a cross section smaller by a factor of $\sim 1.7$ for $m_a=2$ TeV.}.  Both the coherent and incoherent  photon emission are taken into account in the  $\gamma Z$ production cross section.
The latter  and the remaining vector boson fusion contributions  from $ZZ$ and $WW$ channels are also computed with MadGraph5 and are subdominant  when the effective couplings $c_{VV}$ arise from a common source of new physics.

The gluon fusion cross section occurring for a non vanishing $y_t$ includes K-factor corrections up to the next-to-leading order (NLO) in QCD, which have been evaluated by using the model in \cite{Demartin:2014fia}, with MadGraph5\_aMC@NLO \cite{Alwall:2014hca}. We deduce K-factors 
ranging from 2.1 to 1.8 for 0.5 TeV $\lesssim m_a \lesssim$ 2 TeV. Interestingly, we observe the general feature that the production cross section stemming from the top-mediated gluon fusion drops quickly with $m_a$. This is due to the combined effect of the scaling of the gluon PDF at large $x$ and the top-loop function, that vanishes in the limit $m_a/m_t \to \infty$. Another minor effect comes from the running of the strong coupling $\alpha_S$. 

On the contrary, the production cross sections associated with the weak gauge bosons, and especially the photon,  have a much gentler scaling with $m_a$.  This means that the reach with respect to the new pseudoscalar mass $m_a$ is much wider when the new state is produced via the weak gauge boson  rather than via the top-mediated 
gluon fusion.

 \subsection{Production via photon fusion}\label{subsec:photon-fusion}

We will focus now on a very minimal scenario where $a$ does not couple directly to the top, that means setting $y_t=0$ in \eqref{eq:L-eff}. Here the  production relies on purely EW channels, with the dominant contribution coming from photon fusion, as evident from Fig. \ref{fig:xsec}.  
The signal cross section at LO in the narrow-width approximation can be expressed as 
\begin{equation}\label{eq:xsec}
\sigma(p p \to a \to \gamma\gamma) = \frac{8 \pi^2 \,\Gamma(a \to \gamma\,\gamma)}{m_a}\frac{d \mathcal{L^{\gamma\gamma}}}{d m^2_a} \, \text{BR}(a\to \gamma\,\gamma) \ ,
\end{equation}
where $d \mathcal{L^{\gamma\gamma}}/d m^2_a$ denotes the photon luminosity function, which can be extracted from Fig.~\ref{fig:xsec}. From (\ref{eq:rateGammaGamma}) and (\ref{eq:xsec}) it is clear that the $a$ production cross section $\sigma(pp\to a)$ depends quadratically on $c_{AA}$.
In our study of the reach/constraints on diphoton resonances produced by photon fusion we consider a fixed value for BR($a \to \gamma\gamma$), that is BR($a \to \gamma\gamma$)=0.6. This choice is motivated by the specific composite dynamics explanation of the 750 GeV diphoton excess, which will be discussed in section~\ref{sec:model}.  Nevertheless, our results do not depend on this specific choice, since a difference in BR($a \to \gamma\gamma$) will only imply a rescaling of the effective coupling $c_{AA}$. This results in an overall vertical shift of the constraints in the ($m_a$, $c_{AA}$) plane \footnote{Note that this is true under the assumption that dim-6 operators, not included in  the effective Lagrangian (\ref{eq:L-eff}), can be safely neglected. These operators typically give corrections of the order $m^2_a/(4\pi \Lambda_{\rm NP})^2=(c_{AA} m_a/v)^2$, using (\ref{eq:c-lambda}). These corrections become sizable only in a small region of the mass-coupling parameter space. For example, we have, using our previous estimate, corrections $\gtrsim$ 50\% for $m_a\gtrsim 4$ TeV and $c_{AA}\gtrsim 0.045$. }. 

\begin{figure}[]
\begin{center}
\includegraphics[width=0.95\textwidth]{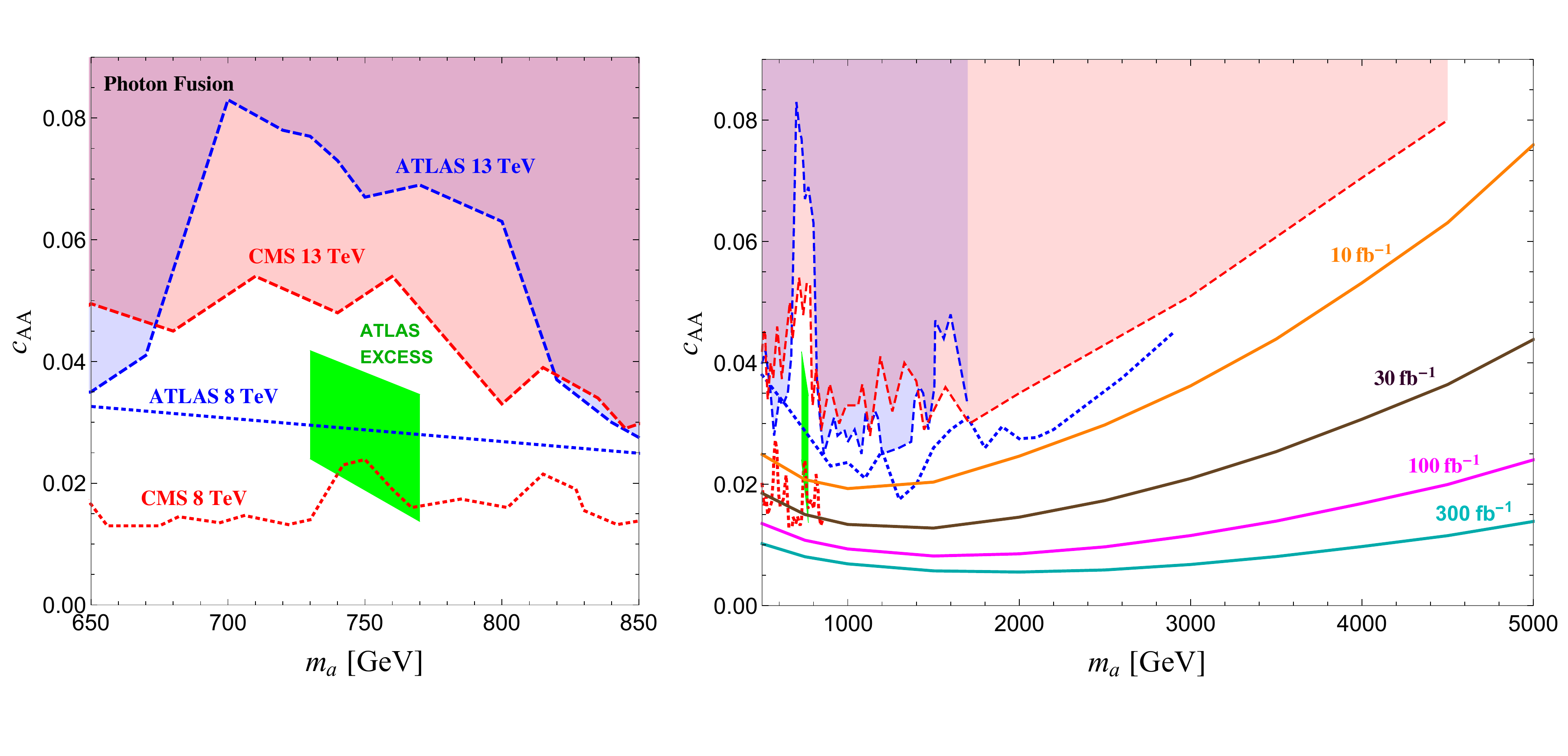} 
\caption{LHC constraints and expected future reach of the diphoton channel in the plane ($m_a$, $c_{AA}$). We assume that $a$ is totally produced by photon fusion and decays to $\gamma\gamma$ with a branching ratio equal to 0.6. {\it Left plot}: ATLAS excess plus LHC-8 and LHC-13 constraints from searches for diphoton resonances in the mass region near 750 GeV. {\it Right plot}:  LHC constraints and expected $2\sigma$ LHC-13 reach for different integrated luminosities: 10, 30, 100, 300 fb$^{-1}$. See the text for details.
 }
\label{fig:aaF}
\end{center}
\end{figure}

We start by examining the region of the effective coupling space able to reproduce the diphoton excess at $m_a \simeq$ 750 GeV. This is shown in the left plot in Fig.~\ref{fig:aaF}, where the green area in the ($m_a$, $c_{AA}$) plane is the region of the parameters fitting the central value of the ATLAS excess within $\pm 1\,\sigma$. We use the  730 and 770 GeV bins  of the ATLAS data \cite{ATLAS}. With the assumed value of BR($a \to \gamma\gamma$) we find that the required $c_{AA} \approx 0.03$ corresponds to a $\Lambda_{ \rm NP} \approx 650$ GeV. This naive estimate suggests the existence of a new physics scale consistent with minimal models of dynamical EW symmetry breaking, that we will explore in the next section.  

We now turn our attention to LHC-8 earlier constraints and confront them with the LHC-13 results.  The dotted and dashed curves in Fig.~\ref{fig:aaF} are respectively the LHC-8 and LHC-13 constraints from searches tailored for diphoton resonances. The curves are obtained using the 95\% C.L. limits given in \cite{ATLAS} (\cite{Aad:2015mna}), which refer to the 13 (8) TeV ATLAS data at 3.2 (20.3) fb$^{-1}$, and in \cite{CMS} (\cite{Khachatryan:2015qba}) for the 13 (8) TeV CMS data with 2.6 (19.7) fb$^{-1}$. We present only the CMS results associated with the narrow-width scenario, which is appropriated for the photon fusion production. 

One observes a slight tension with respect to the 8 TeV data, in particular with those from CMS. Taking into account the uncertainty on the photon fusion production cross section which is about 50\% \cite{Ball:2013hta}, reduces this tension below the 1 $\sigma$ level (see also \cite{Harland-Lang:2016qjy} for a recent discussion). 

 Beyond the excess at 750 GeV it is interesting to explore the  LHC-13 reach of the diphoton channel for larger masses of the new pseudoscalar state. 
We therefore need to estimate the background. As proved in \cite{ATLAS,CMS} a functional form provides a good description of the background. We thus adopt the same estimate of the ATLAS analysis  \cite{ATLAS}, which yields the following best-fit curve for the number of background events as function of the invariant mass $m_{\gamma\gamma}$:
\begin{equation}\label{eq:bckg}
B(x, L) = \frac{L}{3.2\,\text{fb}^{-1}}\left(1-x^{1/3}\right)^{9.9} x^{-2.3}
\end{equation}
with $x=m_{\gamma\gamma}/\sqrt{s}$ and $L$ being the total integrated luminosity. We then estimate the reach by applying the same ATLAS selection criteria in \cite{ATLAS} according to which the transverse energies ($E_T^{\gamma_{1,2}}$) and the pseudo-rapidity ($\eta_\gamma$) of the two leading photons must satisfy
\begin{equation}\label{eq:cuts}
E^{\gamma_1}_T > 0.4 \, m_{\gamma \gamma} \ , \qquad  E^{\gamma_2}_T > 0.3 \,  m_{\gamma \gamma} \ , \qquad |\eta_{\gamma}|<2.37 \qquad ( |\eta_{\gamma}| \notin [1.37,1.52]) \ .
\end{equation}
Furthermore, the photons must be isolated following
\begin{equation}\label{eq:iso}
E^{\rm iso}_T < 0.05\, E^{\gamma_{1,2} }_T + 6 \, \text{GeV} \ ,
\end{equation}
where $E^{\rm iso}_T$ is defined as the transverse energy of the vector sum of all stable particles found within a cone $\Delta R \leq 0.4$ around the photon, neglecting muons and neutrinos. Finally, we assume a 95\% efficiency for the photon identification.
We simulate the signal with  MadGraph5\_aMC@NLO \cite{Alwall:2014hca}, passing the events to PYTHIA \cite{Sjostrand:2006za} for showering and hadronization and to Delphes3 \cite{deFavereau:2013fsa} to mimic detector effects. We use the ATLAS default detector card with the isolation criteria reported in (\ref{eq:iso}). 
After applying the selection cuts above, we obtain signal acceptances from 0.47 for $m_a=$ 0.5 TeV to 0.56 for $m_a=$ 5 TeV. 
The reach is then estimated by assuming a sensitivity $S/\sqrt{S+B}=2$, where $S$ ($B$) represents the number of signal (background) events at a given integrated luminosity, which pass the selection. This gives a conservative estimate of 95\% C.L. limits that LHC-13 will be able to place at different luminosities. The results of this analysis are illustrated in the right plot of Fig. \ref{fig:aaF}. 
As anticipated they can be applied to any model with a pseudoscalar particle entirely produced via photon fusion and where no new decay channels open at higher energies. 
 For example, models featuring an effective photon coupling in the range $0.015 \lesssim c_{AA} \lesssim 0.04$ suggested by the excess, can be ruled out for $a$ masses up to around 2.5, 4  and more than 5 TeV with respectively 10, 30 and 100 fb$^{-1}$ of collected integrated luminosity at the LHC-13. We also observe that the 8 TeV  searches \cite{Aad:2015mna,Khachatryan:2015qba} put so far the strongest constraints, with ATLAS reaching $m_a=2.9$~TeV.
  Intriguingly, the LHC-8 analysis performed by ATLAS is even compatible with another excess around $m_a=$ 1.6 TeV 
 from the LHC-13 run corresponding to a local significance of about 2.8$\sigma$. 
The latter could be excluded in the upcoming run with circa 10 fb$^{-1}$,  shown as the orange curve in the plot.  

Summarizing the results for the photon fusion production mechanism, we have shown that  LHC-13 can test the presence of new pseudoscalar states up to several TeVs. More precise limits can be obtained, following our analysis,  once the photon PDFs will be known more accurately.

 \subsection{Production via top-mediated gluon fusion:  broadening the resonance}\label{sec:GF}
 
 We  now turn on the coupling  $y_t$ to the top, see \eqref{eq:L-eff}, and show that it is possible to fit the 750 GeV diphoton excess while simultaneously accommodating a wide total width $\Gamma_{\rm tot}(a)\approx 45$ GeV ($\Gamma_{\rm tot}(a)/m_a\approx 0.06$), currently preferred by ATLAS.  Indeed, in this case $\Gamma_{\rm tot}(a)$ is dominated by the tree-level decay of $a$ into a $t \bar{t}$ pair, see \eqref{gammattbar}. Of course, the production of $a$ relies on the top-mediated gluon fusion mechanism yielding the signal cross section  
\begin{equation}\label{eq:xsecgg}
\sigma(p p \to a \to \gamma\gamma) = \frac{ \pi^2 \Gamma(a \to g\,g)}{8 \,m_a}\frac{d \mathcal{L}^{g g}}{d m^2_a} \times \text{BR}(a\to \gamma\,\gamma) \ ,
\end{equation}
where $\Gamma(a \to g\,g) \propto y^2_t$ and ${d \mathcal{L}^{g g}}/{d m^2_a}$ is the gluon luminosity function, which can be read from Fig.~\ref{fig:xsec}. Notice that if $\Gamma_{\rm tot}(a)\simeq \Gamma(a \to t\bar{t}) \propto y^2_t$ the cross section (\ref{eq:xsec}) simplifies to
\begin{equation}\label{eq:xsec-cAA}
\sigma(p p \to a \to \gamma\gamma)\simeq  c^2_{AA} \frac{\alpha^2_S |F(m^2_a/4 m^2_t)|^2}{6144 \, \pi}\frac{m^4_a}{m^2_t v^2}\frac{d \mathcal{L}^{g g}}{d m^2_a} \ ,
\end{equation}
it does not depend on  $y_t$ and it is controlled by the effective coupling $c_{AA}$.

\begin{figure}[t!]
\begin{center}
\includegraphics[width=0.95\textwidth]{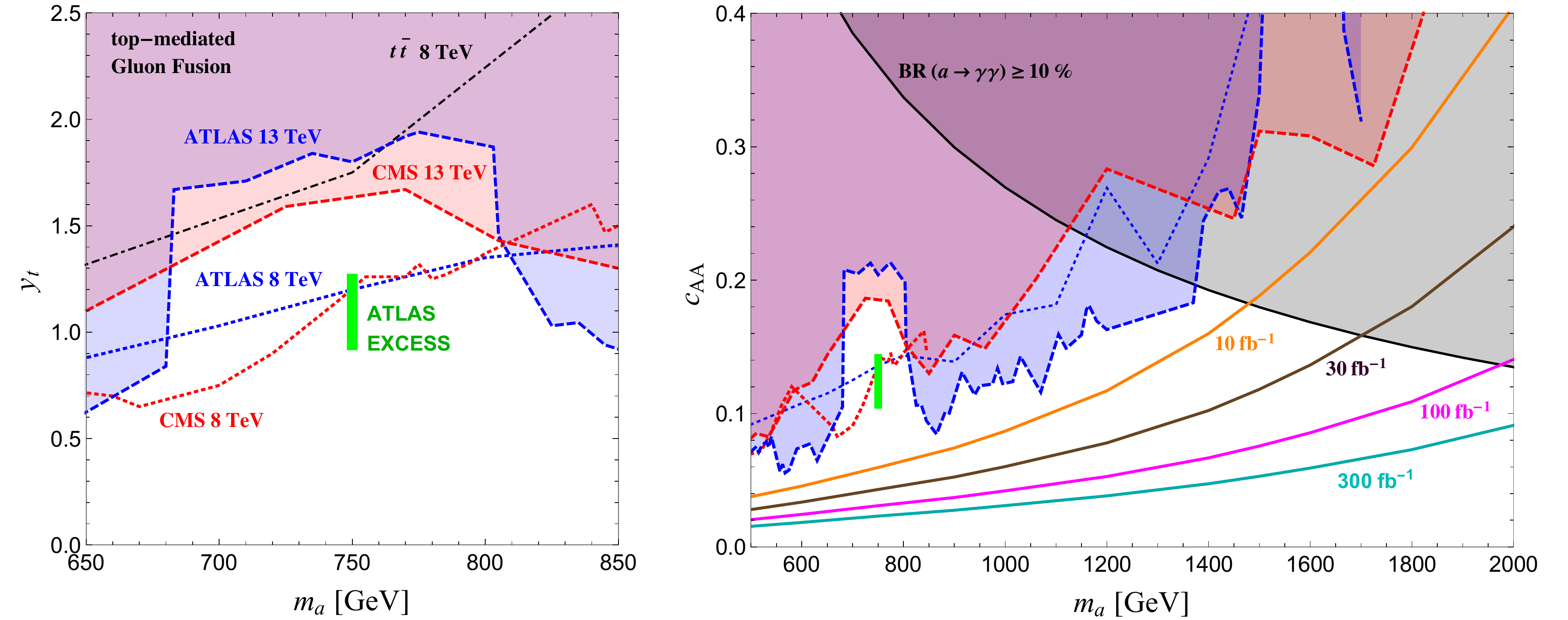} \,~~~~~~~~
\caption{LHC constraints and expected future reach of the diphoton channel in the planes ($m_a$, $y_{t}$) 
({\it left plot}) and  ($m_a$, $c_{AA}$) ({\it right plot}). We assume that $a$ is totally produced by top-mediated gluon fusion. {\it Left plot}: ATLAS excess plus LHC-8 and LHC-13 constraints from searches for diphoton and $t\bar{t}$ resonances in the mass region near 750 GeV. We assume ${\rm BR}(a\to\gamma\gamma)=1\%$. 
{\it Right plot}:  LHC constraints and expected $2\sigma$ LHC-13 reach for different integrated luminosities: 10, 30 ,100, 300 fb$^{-1}$, under the assumption that $\Gamma_{\rm tot}(a) \simeq \Gamma(a \to t\bar{t})$. 
The dark shaded area corresponds to ${\rm BR}(a\to\gamma\gamma)\geq 10\%$ and is indicative 
of a region of the parameter space where relevant corrections to our predictions may apply. See text for details.}
\label{fig:ggF}
\end{center}
\end{figure}

 Using (\ref{eq:xsecgg}) we start by considering  the case of a fixed $\text{BR}(a \to \gamma\gamma)=0.01$ and determine the range of $y_t$ compatible with the 750 GeV excess. The best fit region for a wide-width resonance
  must reproduce simultaneously the two bins in the diphoton invariant mass at 730 and 770 GeV \cite{ATLAS}. This gives the green band shown in the left plot of Fig.~\ref{fig:ggF}, according to which 0.94 $\lesssim y_t \lesssim 1.25$. 
  The blue dashed (dotted) curve in the plot represents the 95\% C.L. limits in \cite{ATLAS} (\cite{Aad:2015mna}) for the 13 (8) TeV ATLAS data with 3.2 (20.3) fb$^{-1}$.  Analogously, the red lines show the CMS bounds
  \cite{CMS} (\cite{Khachatryan:2015qba})  for a wide-width resonance at  13 (8) TeV  and 2.6 (19.7) fb$^{-1}$.
    Part of the parameter space is also excluded by  $t\bar{t}$ resonance searches. 
So far the strongest constraints  on these resonances  are set by  the combined CMS  analyses in different $t\bar{t}$ 
  final states at LHC-8 with 19.7 fb$^{-1}$ \cite{Khachatryan:2015sma}. Such limits are indicated by the black dot-dashed line in the 
  figure~\footnote{More precisely, we use the limits in \cite{Khachatryan:2015sma} for a $Z^\prime$ decaying into $t\bar{t}$ 
  with a  width-over-mass ratio of 0.1. 
  In the case of a pseudoscalar resonance produced via gluon fusion and decaying into $t\bar{t}$ interference effects with the background occur, possibly giving dips in the $t\bar{t}$ invariant mass distribution. This issue has been recently studied in \cite{Djouadi:2016ack} for a resonance at 750 GeV. According to the estimates of the interference effect in \cite{Djouadi:2016ack}, the limits on $y_t$ from the diphoton channel at 8 TeV are still the strongest ones. An experimental analysis for arbitrary resonance masses based on dip-searches does not exist at the moment and the limits we present are conservative. }.
  We also checked that the searches for di-jet resonances \cite{Khachatryan:2015dcf,ATLAS:2015nsi, Aad:2015eha,Khachatryan:2015sja} do  not place any  bound on the relevant parameter space. In contrast to the photon fusion scenario, where the resonance is narrow, now there is no significant tension between the 8 TeV results and the ATLAS diphoton excess. 
  We also verify that for $\text{BR}(a\to \gamma\,\gamma)=0.01$ the values of $y_t$ that reproduce the diphoton excess imply a total width in the range 35 GeV $\lesssim \Gamma_{\rm tot}(a) \lesssim$ 62 GeV. On the other hand, taking $y_t=1$, the same range of $\Gamma_{\rm tot}(a)$ is obtained for $\text{BR}(a\to \gamma\,\gamma)$ between 0.009 and 0.016. Henceforth, we conclude that only models predicting $y_t \approx 1$ and $\text{BR}(a\to \gamma\,\gamma)\approx 0.01$ can reproduce the ATLAS best-fit of the diphoton invariant mass. \\

 Under the assumption that the total width is dominated by the $t\bar{t}$ decay rate, as argued before, the cross section does not depend on $y_t$.  We can therefore use \eqref{eq:xsec-cAA} to study the LHC-13 reach/constraints of the diphoton channel on the $(m_a , c_{AA})$ plane, as done for the photon fusion production.
 We estimate the background according to (\ref{eq:bckg}) and we apply the cuts in (\ref{eq:cuts}, \ref{eq:iso}) to the signal events simulated with Madgraph5+PYTHIA+Delphes3. Note that the gluon fusion production includes the NLO K-factors discussed in section~\ref{sec:xsec}. We obtain signal acceptances from 0.43 at $m_a=0.5$ TeV to 0.46 at 2 TeV.  
The limits we obtain from this analysis are shown on the right plot of Fig.~\ref{fig:ggF}.  
We observe that the values of $c_{AA}$ that can explain the ATLAS excess at 750 GeV lay in the interval 
$0.10\lesssim c_{AA}\lesssim 0.14$. These are larger than in the photon fusion case by a factor around 3. Any model with values of $c_{AA}$ in this range and for which $\Gamma_{\rm tot}(a) \simeq \Gamma(a \to t\bar{t})$ can be ruled out up to $m_a=$ 1.3, 1.6 and 2 TeV with respectively  10, 30, and  100  fb$^{-1}$ at LHC-13. 
Furthermore, we observe that, in contrast to the photon fusion scenario shown in Fig. \ref{fig:aaF}, the current strongest constraints are now set by the 13 TeV ATLAS data rather than the LHC-8 results. This depends on the larger 
increase of the gluon PDF, compared to the photon PDF, passing from $\sqrt{s}=8$~TeV to $\sqrt{s}=13$~TeV.

Finally, the gray-shaded area in the plot indicates the region of the parameter space  where the $\text{BR}(a \to \gamma\gamma)$ is larger than $10\%$ (calculated for $y_t=1$).  Above this value, corrections to the approximated expression for the cross section in \eqref{eq:xsec-cAA} become relevant and we further expect non negligible contributions to the production cross section from photon fusion. \\
 
We briefly summarize   the salient results presented in this section. We have learned that: i) it is possible to  directly photo-produce the 750 GeV excess   without the need of any colored state; ii) in this case the resonance is narrow; iii) beyond the 750 GeV excess LHC-13 will be able to constrain new resonances up to 5 TeV with 100 fb$^{-1}$; iv) a broad width for the 750 GeV excess can be minimally achieved when the new resonance couples to the SM top with a perturbative Yukawa-like coupling; v) the reach in this case is around 2 TeV for 100 fb$^{-1}$. 

We now move to a concrete realization of these scenarios in terms of minimal models of composite dynamics. 
   
\section{Diphoton resonances in scenarios of minimal composite dynamics}\label{sec:model}

We consider now a UV completion of the effective field theory introduced in section~\ref{sec:xsec}. The theory naturally
predicts a new pseudoscalar  particle in the TeV range that can be revealed by LHC searches for diphoton resonances. 
This is realized in a minimal framework of composite dynamics naturally addressing the SM hierarchy 
problem. The model includes  $N_F$ new fermions, hereafter dubbed techniquarks, which engage  in a new (asymptotically free) $SU(N_T) $ gauge interaction. 
For massless techniquarks transforming in a complex representation of the underlying gauge composite theory, and in absence of EW interactions, the theory preserves an $SU(N_F)_L\times SU(N_F)_R$  global chiral symmetry which spontaneously breaks around  $\Lambda_T \gtrsim 1$ TeV to the  custodial group $SU(N_F)_V$. 
A larger global quantum symmetry occurs when the underlying fermions belong to a (pseudo)real representation of the fundamental composite gauge group, i.e. $SU(2N_F)$.  Enlarged global symmetries are interesting since they lead to composite pseudo-Goldstone Higgs realizations  \cite{Kaplan:1983fs, Kaplan:1983sm}.

The $N_F^2-1$ Goldstone bosons which arise from the breaking of the axial-vector symmetry are, therefore,  massless  composite pseudoscalar fields made up of the new fermions and their antiparticles. It is customary to describe them, as well as possible new pseudoscalar isosinglets of flavor, by a $N_F\times N_F$ unitary matrix $\mathcal{U}$, which 
transforms bilinearly under a chiral rotation:
\begin{equation}
	\mathcal{U} \to u_L\,\mathcal{U} \, u_R^\dagger \label{trU}
\end{equation}
with $u_{L/R} \in SU(N_F)_{L/R}$. When the EW $SU(2)_W\times U(1)_Y$ gauge interactions are switched on, three of the Goldstone bosons become the  longitudinal
degrees of freedom of the $W^\pm$ and $Z$ gauge bosons, while the photon remains massless.

 In the following we consider for simplicity an underlying theory with $N_F=2$ 
techniquarks $U$ and $D$, which transform under a given complex  representation $R$ of the new  gauge group. Furthermore,  within this minimal composite scenario, we assume that
$U$ and $D$ do not carry SM color, i.e. they are singlet of $SU(3)_C$. On the other hand, the left-handed and right-handed chiral projections of the new fundamental fermions have non-trivial quantum numbers under  the EW gauge symmetry.
In order to cancel Witten  \cite{Witten:1982fp} and gauge anomalies we further enlarge the fermion sector with new leptons, $N$ and $E$, with proper weak gauge and hypercharge quantum numbers \cite{Foadi:2007ue}.
In particular,  the left-handed projections $Q_L\equiv(U_L, D_L)$ and $L_L\equiv(N_L, E_L)$ transform in the fundamental of $SU(2)_W$, while the corresponding right-handed fields $U_R$, $D_R$, $N_R$ and $E_R$ 
are weak isosinglets. The hypercharge assignments which make the theory anomaly free are
\begin{align}
\begin{split}\label{HYtechniquarks}
	& Y(Q_L) \,=\, \frac{y}{2}\,, \quad\quad Y(U_R/D_R)\,=\,\frac{y\pm1}{2}\,, \\ 
	&  Y(L_L)\,=\, -d(R)\,\frac y 2\,\quad\quad Y(N_R/E_R)\,=\,\frac{-d(R)\,y\pm 1}{2}  \, ,
	\end{split}
\end{align}
where $y$ is a real parameter and $d(R)$ denotes the dimension of the techniquark representation. The electric charge operator is defined as $Q=T_3+Y$, where $T_3$ is the weak isospin generator. 

The condensation of the techniquarks, i.e. $\langle 0| \overline{U} U +\overline{D} D|0\rangle \neq 0$,
induces the chiral symmetry breaking pattern $SU(2)_L\times SU(2)_R \to SU(2)_V$, as well as the correct breaking of the EW symmetry. The latter is embedded by gauging a subgroup 
of $SU(2)_L\times SU(2)_R \times U(1)_V$. The spectrum of the massive states depends sensitively on the specific underlying dynamics. In QCD-like theories we expect new resonances to appear at energy scales around $\Lambda_T$ as further confirmed by recent lattice simulations \cite{Lewis:2011zb,Hietanen:2014xca} \footnote{ Recently also the vector decay constants \cite{Arthur:2016dir} have been computed for the minimal composite template that can be used for technicolor, composite Goldstone Higgs, and even models of strongly interacting massive particles for dark matter of either asymmetric \cite{Nussinov:1985xr,Barr:1990ca,Gudnason:2006ug,Gudnason:2006yj,Ryttov:2008xe,Frandsen:2009mi,Frandsen:2011kt} or mixed nature \cite{Belyaev:2010kp}, or that can reach the observed relic density via three to two number changing interactions \cite{Hochberg:2014kqa,Hansen:2015yaa}.}.  If, however, the dynamics is not QCD-like the spectrum can be much more compressed   \cite{Dietrich:2005jn,Foadi:2012bb}  and furthermore the top-interactions can, via quantum effects, further reduce the lightest scalar mass \cite{Foadi:2012bb}. Analytical \cite{Sannino:2004qp,Dietrich:2006cm,Bergner:2015dya} and numerical efforts \cite{Catterall:2007yx,Hietanen:2009zz,DelDebbio:2010hx,DeGrand:2011qd,Appelquist:2011dp,DeGrand:2010na,Fodor:2015zna,Hasenfratz:2015ssa,Athenodorou:2014eua} have been dedicated to determine whether fermionic gauge theories display large distance conformality and investigate their spectrum. For the sextet composite model \cite{Sannino:2004qp,Dietrich:2005jn,Dietrich:2005wk}, lattice results \cite{Fodor:2012ty,Fodor:2015vwa,Fodor:2016wal} suggest that the theory is either very near-conformal or conformal. In the latter case  interactions responsible for giving masses to the SM fermions can modify the conformal-boundary inducing an ideal near-conformal  behaviour \cite{Fukano:2010yv}. Furthermore, near-conformality alleviates tension with EW precision measurements \cite{Appelquist:1998xf} and flavor changing neutral current constraints \cite{Holdom:1983kw}.

An interesting state is the pseudoscalar $a$ associated with the $U(1)_A$ axial anomaly of the underlying gauge theory, which is analogous to the $\eta^\prime$ of QCD. This is the one we assume to induce the diphoton excess and it is included as a singlet state in the matrix $\mathcal{U}$. 
The pseudoscalar degrees of freedom  are therefore parametrized via the following unitary matrix $\mathcal{U}$ transforming as in \eqref{trU} for $N_F=2$, that is \footnote{ We assume the large-$N_T$ relation between the decay constant of the singlet $a$ and the technipions, namely $F_a = F_T \left(1 + \mathcal{O}(1/N_T) \right). $}
\begin{equation}
	\mathcal{U}\; = \; e^{i \Phi /F_T}\;=\; \exp\left[ \frac{i}{F_T}\left(a\,+\, {\bm{\tau}}\cdot {\bm{\Pi}}\right)  \right]\,,
\end{equation}
where $\bm{\tau}\equiv\left(\tau_1,\,\tau_2\,,\tau_3 \right)$ are the standard Pauli matrices.  The technipions $\bm{\Pi}\equiv(\Pi^1,\Pi^2,\Pi^3)$ couple with strength $F_T$ to the 
axial-currents corresponding to the broken generators of the chiral symmetry. The linear combinations $\Pi^\pm\equiv\left(\tau_1\mp i \tau_2\right)/\sqrt 2$ and $\Pi^0\equiv \Pi_3$ become the longitudinal polarizations of the   $W^\pm$ and the $Z$ bosons, respectively, which thus acquire masses
\begin{equation}
	m_W^2 \;=\; \frac 1 2\,g_W\, F_T^2\,,\quad \quad m_Z^2\;=\;\frac 1 2\sqrt{g_W^2+g_Y^2} F_T^2\,,\label{mgauge}
\end{equation}
with $F_T=v=246$ GeV. 

Heavy vector mesons are also generated by the composite dynamics. The relevant states  here are given by one isosinglet ($\omega$) and one isotriplet ($\bm{\rho}$) spin-1 resonance, which is described by the matrix
\begin{equation}
	\mathcal{V}^\mu \; = \; \lambda\,\left( \omega^\mu\,+\,\bm{\tau}\cdot\bm{\rho}^\mu \right)\,,
\end{equation}
where $\lambda$ is a parameter connected to the $\rho$-$\Pi$-$\Pi$ effective coupling. These new vector mesons have typically masses of the order of a few TeVs \cite{Lewis:2011zb,Hietanen:2014xca,Arthur:2016dir}. In this case,  their mixing with the technipions, namely  the EW gauge bosons, which could in principle have an impact on the decays of $a$, can be safely neglected.
Furthermore, since this study will be focused on the phenomenology of the pseudoscalar meson, we will assume in the following that the vectors are decoupled. 
A specific study of collider signatures of composite vector mesons that might be lighter because of near-conformal dynamics is very interesting and will be presented elsewhere\footnote{As for the precision observables, we comment here that the most naive estimate for the $S$ parameter yields $S\approx \frac{N_T}{6 \pi}$. This estimate must be taken with the {\it grain of salt} and it would prefer smaller values of $N_T$. For example, for $N_T=4$, $S$ is of the order of 0.2. If a near-conformal dynamics would be present, it would alleviate the tension with experimental data \cite{Appelquist:1998xf}. }.
The low energy physics below the composite scale $\Lambda_T$ is  encoded in the following effective Lagrangian $\mathcal{L}_{\rm eff}$ constructed to respect all the symmetries of the underlying theory. 
Neglecting gauge fixing  and Faddeev-Popov terms, the skeleton of the effective Lagrangian reads:
\begin{equation}
	\mathcal{L}_{\rm eff}\; =\; -\,\frac 1 4 B^{\mu\nu} B_{\mu\nu}\,-\,\frac 1 4 {\rm Tr}\left[W^{\mu\nu} W_{\mu\nu}\right]\,+\,\mathcal{L}_{\rm comp}\,+\,\mathcal{L}_{\rm ferm}\label{LagrComp}\,,
\end{equation}
where $B^{\mu\nu}$ and $W^{\mu\nu}$ are the field strengths of the EW gauge bosons. 
Here the techniquarks and the $SU(N_T)$ gauge fields are integrated out, while the SM fermions and the new leptons $N$ and $E$ are coupled to the composite states
via $SU(2)_W\times U(1)_Y$ invariant operators in $\mathcal{L}_{\rm ferm}$, which originates from an unspecified extended gauge dynamics (EGD). The details of the underlying
theory are encoded in the effective parameters of $\mathcal{L}_{\rm eff}$. The measurement of some of them in 
current and future collider experiments  will allow to infer more insights about the new strong dynamics that can be confronted with first principle lattice predictions. Here we 
point out the importance of searches for diphoton resonances. 
In fact, in our theory the composite meson $a$ can be naturally identified with the particle responsible for local excess in the diphoton invariant mass at 750 GeV.

The Lagrangian $\mathcal{L}_{\rm comp}$ in (\ref{LagrComp}) contains  the gauge invariant effective interactions involving  $\mathcal{U}$. In particular, the leading terms are: 
\begin{equation}
	\mathcal{L}_{\rm comp} = \frac{1}{4}F_T^2\,{\rm Tr}\left[\left(\mathcal{D}^\mu \mathcal{U}\right)^\dagger\,\mathcal{D}_\mu\mathcal{U} \right]\,+\,\mathcal{L}_{m_a}\,+\,\mathcal{L}_{\rm WZW}\,+\,\dots\label{LcompletaT}
\end{equation}
Consistently with the effective theory approach outlined in the previous section, we neglect in $\mathcal{L}_{\rm comp}$ terms of the order $p^6$, which may be relevant at energies $p\sim 4\pi v \approx 3$ TeV. 

The covariant derivative in \eqref{LcompletaT} takes the standard form
\begin{equation}
	\mathcal{D}^\mu\mathcal{U}\;=\;\partial^\mu\mathcal{U}\,-\,i\, A^\mu_L\mathcal{U}\,+\,i\,\mathcal{U} A^\mu_R\,,
\end{equation}
with
\begin{eqnarray}
 A^\mu_L &=&   g_Y \left(Q-\frac 12 \tau_3 \right)\,B^\mu\,+\, \frac 12\, g_W\, \bm{\tau}\cdot\bm{W}^\mu\,, \quad\quad  A^\mu_R \;=\;  g_Y \,Q\,  B^\mu\,, \label{AR}
\end{eqnarray}
where $Q$ is the electric charge matrix of the fundamental techniquarks. Under $SU(2)_W\times U(1)_Y$ gauge transformations
\begin{eqnarray}
	A^\mu_L & \to & u_L \,A^\mu_L \,u_L^\dagger\,-\,i\,\partial_\mu u_L\, u_L^\dagger\,,\quad\quad	 A^\mu_R \; \to \; u_R A^\mu_R\, u_R^\dagger\,-\,i\,\partial_\mu u_R\, u_R^\dagger\,.
\end{eqnarray}
where $u_L\in SU(2)_W$ and $u_R\equiv \exp(i\, \theta(x) \,\tau_3/2)$. Notice that the matrix $\mathcal{U}$ is not sensitive to the hypercharge parameter $y$, because it is formed by
pairs made by a techniquark and  the corresponding antiparticle, see \eqref{HYtechniquarks}.
 
The physical masses of the EW gauge bosons in \eqref{mgauge} arise directly from the
first operator in \eqref{LcompletaT}, whereas   $\mathcal{L}_{m_a}$ provides a mass term for $a$. Taking the techniquarks in the fundamental representation,
$d(R=\text{Fund})=N_T$, the corresponding gauge invariant Lagrangian is
\begin{equation}
	\mathcal{L}_{m_{a}} \;= \; \frac{1}{32}\,m_a^2\,F_T^2\,{\rm Tr}\big[ \ln \mathcal{U} -\ln \mathcal{U}^\dagger\big]^2\,,
\end{equation}
where the mass $m_a$ is given at leading order in the large $N_T$  limit by the Witten-Veneziano relation \cite{Witten:1979vv,Veneziano:1979ec}
\begin{equation}\label{mamass}
	m_a \; = \; \sqrt{\frac{2}{3}}\frac{F_T}{f_\pi}\,\frac{3}{N_T}\,m_{\eta_0}\,\approx\, \frac{6}{N_T}~\text{TeV}\,.
\end{equation}
Here $f_\pi=92$ MeV is the pion decay constant and $m_{\eta_0}=849$~MeV. $\eta_0$ indicates the QCD SU(3) flavor singlet state in the chiral limit with 
	$m_{\eta_0}^2 = m_{\eta^{\prime}}^2\,+\,m_\eta^2\,-\,2\,m_K^2$. As we can see from  (\ref{mamass}),  for techniquarks in the fundamental  representation of the new gauge interaction
we naturally expect values of the pseudoscalar mass in the TeV range. This is actually also the case for different representations  \cite{Sannino:2003xe,DiVecchia:2013swa}, for which expression \eqref{mamass} cannot be applied. 
In our numerical analysis we will fix $d(R)=6$.

\begin{figure}[t!]
\begin{center}
\includegraphics[width=0.45\textwidth]{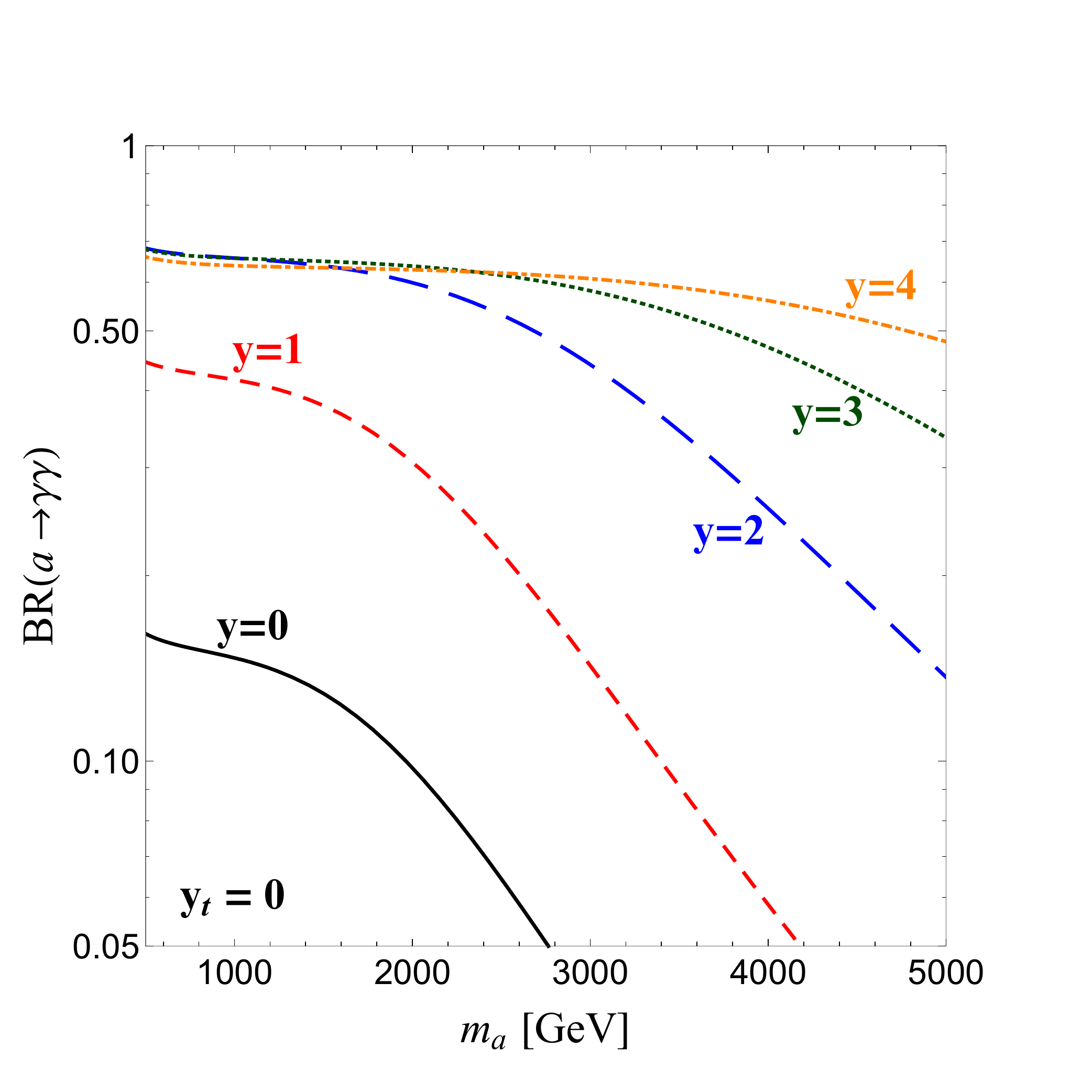} \;\;\;\;
\includegraphics[width=0.45\textwidth]{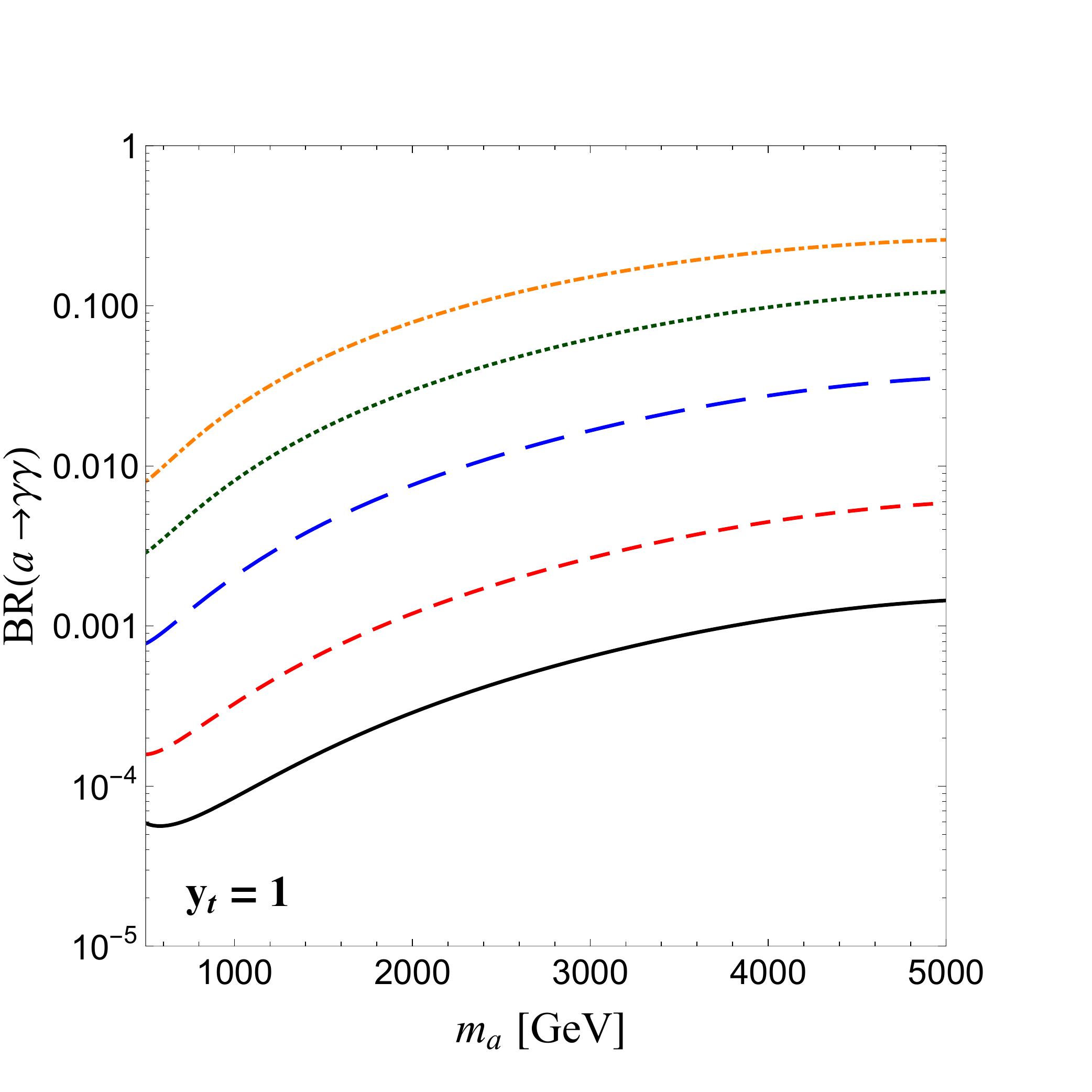}
\caption{Branching ratio  into two photons of the pseudoscalar $a$ as a function of its mass for several values of the hypercharge parameter $y$. The scenario without (with) a direct coupling to the 
top quark, $y_t=0$ ($y_t=1$), corresponds to a narrow (wide) resonance for $m_a=750$ GeV. See the text for details.}
\label{fig:1}
\end{center}
\end{figure}

The  gauged Wess-Zumino-Witten Lagrangian $\mathcal{L_{\rm WZW}}$ is 
a topological term stemming from chiral anomalies associated with the global axial-vector currents.
The complete expression of such term is reported in  (\ref{GWZW}) and  parametrizes EW processes which can be directly tested at the LHC. 
In particular, we are interested here in  the production mechanisms of the $a$ resonance at LHC  and its decays into EW gauge bosons via the relevant topological terms.
We have  at leading order in the derivative expansion of the theory
\bea
\mathcal{L}_{\rm WZW} &=& -\, \frac{i 5C}{F_T}\epsilon_{\mu\nu\rho\sigma}{\rm Tr}\big[\Phi\big(\partial^\mu A_L^\nu\partial^\rho A_L^\sigma+\partial^\mu A_R^\nu\partial^\rho A_R^\sigma
 \,+\,\partial^\mu\left(A_L^\nu+A_R^\nu \right)\partial^\rho\left(A_L^\sigma+A_R^\sigma \right) \big)\big] \nonumber\\
&& +\, \frac{5C}{F_T^3}\epsilon_{\mu\nu\rho\sigma}{\rm Tr}\big[\partial^\mu\Phi\partial^\nu\Phi\partial^\rho\Phi\left(A_L^\sigma+A_R^\sigma\right)\big]+\ldots\,,
\label{WZW}
\eea
with  $C=-i d(R)/(240\,\pi^2)$. 
The matching between (\ref{WZW}) and the effective Lagrangian in (\ref{eq:L-eff}) gives the following 
effective couplings:
\begin{align}
\begin{split}
	c_{AA} =& \left(1+y^2\right)\, e^2\,\frac{d(R)}{8\,\pi^2}\,, \quad\quad \quad\quad c_{AZ} \;=\; \frac{1-2(1+y^2)s_W^2}{2\,c_W\,s_W}\, e^2\,\frac{d(R)}{8\,\pi^2}\,,\\
	c_{ZZ} =& e^2\,\frac{1-3s_W^2+3(1+y^2)s_W^4}{3\,c_W^2\,s_W^2}\frac{d(R)}{8\,\pi^2}\,,\quad\quad c_{WW} \;=\; e^2\,\frac{1}{s_W^2}\frac{d(R)}{24\,\pi^2}\,.       \label{effc}    
\end{split}
\end{align}
  The second term in   (\ref{WZW}) mediates the three-body decay process  $a\to \Pi~\Pi~V$, where $V$ is one EW transverse gauge boson, $V=\gamma /Z/W^\pm$, and $\Pi$ are the longitudinal components of $Z/W^\pm$. In the limit $m_{\Pi^\pm}\approx m_{\Pi^0}\equiv m_\Pi$, the three-body  partial decay rate of $a$  reads
\begin{eqnarray}
 	\Gamma(a \to \Pi~\Pi~V)&=& \frac{m_a^3}{122880\,\pi^3}\,\frac{m_a^4}{F_T^6}\Bigg[\sqrt{1-4u^2}\Big(1-2u^2\left(14+47u^2-80u^4+60u^6\right)\Big)\nonumber\\
					&& +\,240 u^4\left(-1+2u^2-3u^4+2u^6\right)\ln\left(\frac{2 u}{1+\sqrt{1-4u^2}}\right)\Bigg]\,c_{\Pi\Pi V}^2\,, \label{3dec}
 \end{eqnarray}
where $u\equiv m_\Pi/m_a$ and
\begin{eqnarray}
&&	c_{\Pi^+\Pi^- \gamma}  =  e\,\frac{d(R)}{12\,\pi^2}\,,\quad\quad	c_{\Pi^+\Pi^-  Z} \; =\;  \frac{1-2s_W^2}{2c_W s_W}\,c_{\Pi^+\Pi^-  \gamma}\,,\quad\quad  c_{\Pi^\pm\Pi^0 W^\pm} \;=\; \pm \,
 \frac{1}{2 s_W}\,c_{\Pi^+\Pi^- \gamma}\,.
\end{eqnarray}
 
 \subsection{WZW induced photon fusion production}
 
 Having spelled out the underlying dynamics and provided the associated effective Lagrangian, we can now turn to the diphoton process. We first analyse the most minimal case in which the gauged WZW term is  simultaneously responsible for the production of $a$ and its decay into photons. This allows us to provide critical information on the decay rates (and production) of $a$ in several related channels when compared to the blind effective approach used in the previous section.  In  table~\ref{tab:1} we report  the total decay rate $\Gamma_{\rm tot}(a)$ and the partial widths of $a$ into EW gauge bosons for $m_a=750$ GeV and different choices of the techniquark parameter $y$. Notice that for this value of the mass the 
 diphoton resonance is quite narrow, which seems to be preferred by the CMS results \cite{CMS}. The diphoton decay channel dominates the total width for  a non-zero value of the parameter $y$ entering the hypercharge assignment \eqref{HYtechniquarks}.  These results have been obtained in \cite{Molinaro:2015cwg} with the important difference that in that work the production mechanism was induced by a new vector-like colored fermion with a mass around one TeV.   
 
 From the table we observe that there is a strong suppression in the production of one photon and one $Z$ for $y\approx1$, due to the cancellation in  the expression of the effective coupling $c_{AZ}$. 
 As for the $ZZ$ and $WW$ channels, they are subdominant for $y>1$. On the other hand, the three-body decay from \eqref{3dec}, which is negligible for $m_a=750$ GeV, may have a relevant contribution in the total decay rate for larger masses, i.e.  $m_a\gtrsim 1.5$ TeV, because the corresponding partial decay width increases $ \propto m_a^7$. This is manifest in the left panel of Fig.~\ref{fig:1}, which shows the branching ratio of $a$ decaying into two photons versus  $m_a$, for fixed values of $y$. In the case of $y\gtrsim 2$ and $m_a \lesssim 2$ TeV the diphoton decay rate is anyway the largest one.

\begin{figure}[t!]
\begin{center}
\includegraphics[width=0.5\textwidth]{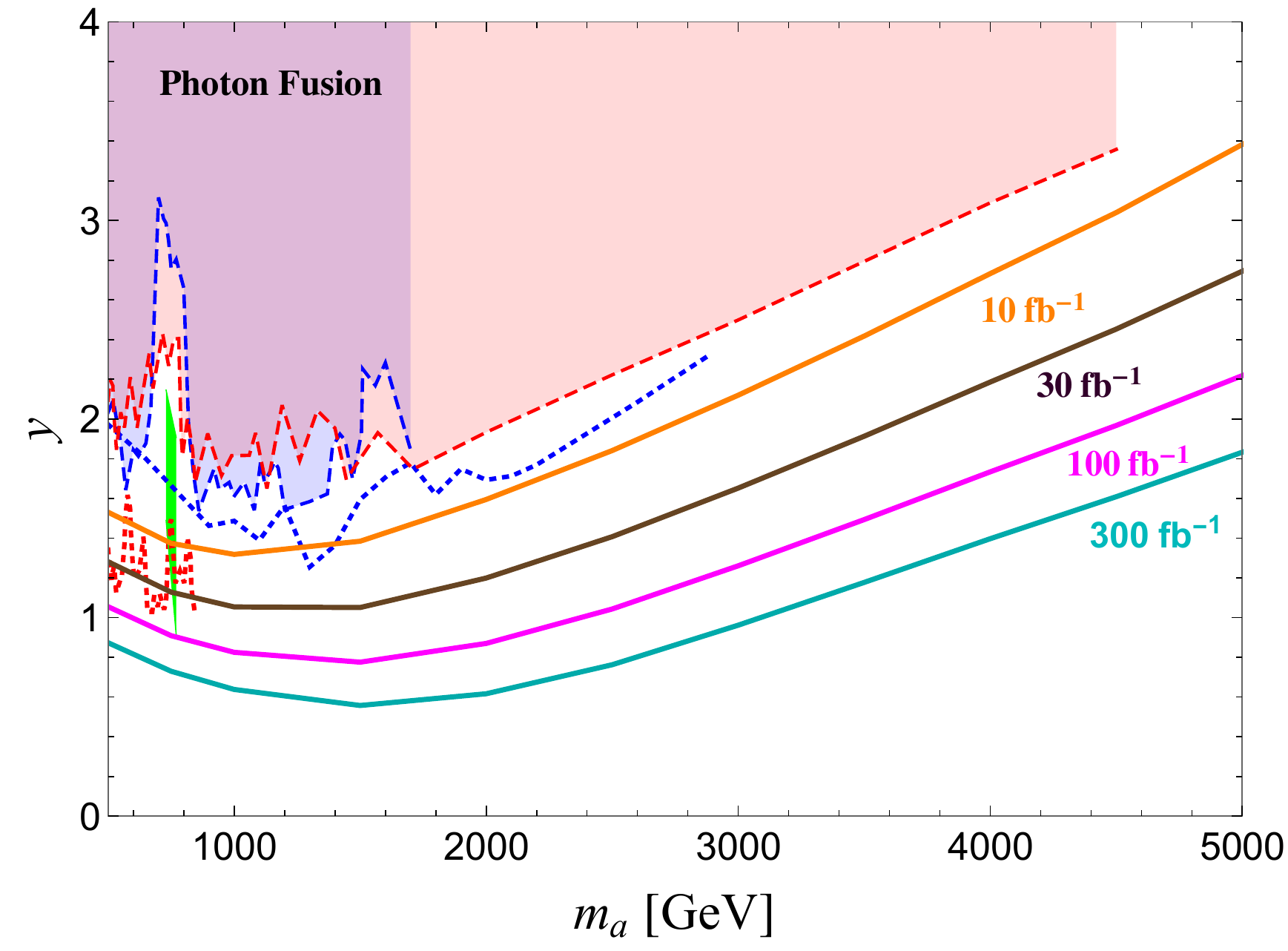}
\caption{LHC constraints, 750 GeV ATLAS excess and expected LHC-13 future reach of the diphoton channel  in the plane ($m_a$, $y$). Here $a$ is totally produced by photon fusion. We fix the dimension of the techniquark representation to $d(R)=6$. The plot legend is the same as in Fig. \ref{fig:aaF}.}
 \label{fig:aaF-EtaPrime}
\end{center}
\end{figure}

The constraints on the effective coupling to the photons $c_{AA}$ as well as the values of the parameters
explaining the ATLAS excess, reported in Fig.~\ref{fig:aaF}, can be directly interpreted in the plane ($m_a$, $y$) for 
$m_a\lesssim 2$ TeV, where the effect of the three-body decays is negligible and ${\rm BR}(a\to\gamma\gamma)$
is independent of $m_a$, see (\ref{eq:rateGammaGamma}-\ref{eq:rateWW}) and the left plot in Fig.~\ref{fig:1}.
Our results are shown in Fig.~\ref{fig:aaF-EtaPrime}, where the dashed (dotted) curves 
represent the ATLAS and CMS limits at 13 (8) TeV as in Fig.~\ref{fig:aaF}, and 
 the green area indicates 
 the interval $0.9\lesssim y \lesssim 2.1$ which fits  the ATLAS data at $m_a=750$ GeV. For this range of the hypercharge one finds $0.4 \lesssim {\rm BR}(a\to\gamma\gamma)\lesssim 0.7$.
  Hence, we compute the LHC-13 future constraints on $y$, following the analysis done in 
 subsection~\ref{subsec:photon-fusion}, where we include  the variation in the diphoton branching ratio 
 due to the three-body decay channels. In conclusion, we find that it is possible to probe values of $y$ as small as $0.6$, for $m_a \lesssim 2$ TeV,  with a luminosity of 300 fb$^{-1}$ at LHC-13. Notice that for resonance masses above about 3 TeV our results can only be taken as a rough estimate because, on top of the uncertainties on the photon PDFs, $\mathcal{O}(p^6)$ terms not included in the chiral Lagrangian in (\ref{LcompletaT}) might be relevant. Given these caveats, we still think that an $\mathcal{O}(1)$ estimate at these energies provides a useful guidance to experiments.  \\

\begin{table}[t!]
{\centering
\catcode`?=\active \def?{\hphantom{0}}
\begin{tabular}{!{\vrule width 1pt}@{\quad}>{\rule[-2mm]{0pt}{6mm}}l@{\quad}!{\vrule width 1pt}@{\quad\quad}c@{\quad\quad}|@{\quad\quad}c@{\quad\quad}|@{\quad\quad}c@{\quad\quad}|@{\quad\quad}c@{\quad\quad}|@{\quad\quad}c@{\quad\quad}!{\vrule width 1pt}}
\Xhline{2\arrayrulewidth}
 \Xhline{3\arrayrulewidth}
  $y$ &  0  &  1 & 2 &  3 & 4   \\[0.2mm]
  \hline
 $\Gamma_\text{tot}(a)$  [MeV]   		   &  13    &  18        & 73    &  290    &  870  \\[0.2mm]
  \hline
 $\Gamma(a\to \gamma\gamma)$  [MeV] 	   &  1.9   &  7.7      & 48    &  190  & 560  \\[0.2mm]
 \hline
 $\Gamma(a\to \gamma Z)$  [MeV]		   &  1.5   &  0.030  & 8.9   &  68    & 240  \\[0.2mm]
 \hline
 $\Gamma(a\to ZZ)$  [MeV]  			   &  1.4   &  2.4      & 7.7   &  23    & 57    \\[0.2mm]
   \hline
 $\Gamma(a\to W^+W^-)$  [MeV]   		   &   \multicolumn{5}{c@{\quad}!{\vrule width 1pt}}{7.5}  \\[0.2mm] 
 \hline
 $\Gamma(a\to \text{3-body})$  [MeV] 	   &  \multicolumn{5}{c@{\quad}!{\vrule width 1pt}}{0.10}  \\[0.2mm] 
\Xhline{3\arrayrulewidth}
\end{tabular}}
\caption{\label{tab:1} Total and partial decay widths from the WZW term in (\ref{WZW}) for a narrow pseudoscalar resonance of mass $m_a=750$ GeV as a function of the hypercharge of the  fundamental techniquarks. We fix the dimension of the techniquark representation to $d(R)=6$. }
\end{table}
%

 \subsection{Top-mediated gluon fusion and WZW induced diphoton decay}
 
Now we turn to a more general realization in which we allow for  direct couplings of $a$ to the SM fermions.
These may be originated by the same EGD responsible for the generations of the fermion masses.
Independently of the underlying theory, such interactions are parametrized at low energy by  gauge invariant effective operators included in $\mathcal{L}_{\rm ferm}$ of \eqref{LagrComp}.
We expect, as for the Higgs boson $h$, the strongest interactions to arise for the third quark generation. Indeed, indicating the top and bottom quark mass eigenstates with $Q_3\equiv (t, b)^T$, their embedding in  $\mathcal{L}_{\rm ferm}$ is given by 
the $SU(2)_W\times U(1)_Y$ invariant effective operators 
\begin{eqnarray}
	 \mathcal{L}_{\rm ferm}  & =& i\, \overline{Q}_{3 L}\gamma_\mu \left(\partial^\mu\,-\,i A_L^\mu\right) Q_{3L}+i \,\overline{Q}_{3 R}\gamma_\mu \left(\partial^\mu\,-\,i A_R^\mu\right)  Q_{3R}\nonumber\\
	&&\,-\,Y_1\, F_T\, f_1(h)\,\left(\overline{Q}_{3 L} {\mathcal U} Q_{3R}\,+\,\overline{Q}_{3 R} {\mathcal U}^\dagger Q_{3L} \right) 
	 \,-\,Y_2\, F_T\, f_2(h)\,\left(\overline{Q}_{3 L} {\mathcal U}\tau_3\, Q_{3R}\,+\,\overline{Q}_{3 R} \tau_3\, {\mathcal U}^\dagger Q_{3L} \right)+\ldots\label{YukawaU}
\end{eqnarray}
where $Q_{3L/R}=P_{L/R}\, Q_3$ and $P_{L/R}=(\bm{1}\mp\gamma_5)/2$ are the  chiral projectors. At leading order in $h/F_T$ the functions $f_{1,2}(h)$ are 
\begin{equation}
	f_{1,2}(h)\;=\; 1\,+\,c_{1,2}\,\frac{h}{F_T}\,+\,\ldots \end{equation}
The coefficients $c_{1,2}$ are constrained by measurements of the 
 Higgs couplings to the SM fermions \cite{Agashe:2014kda}, whereas  $Y_{1,2}$ reproduce  the top and bottom masses:
 $m_t=(Y_1+Y_2)\, F_T$ and $m_b=(Y_1-Y_2)\,F_T$ \footnote{We assume that the Yukawa couplings $Y_{1,2}$ are generated by an Extended Technicolor (ETC) dynamics. In the case of the top, for example, one expects \cite{Chivukula:1987py,Randall:1992vq}
 \[ 
 m_t \approx \frac{g^2_{ETC}}{M^2_{ETC}} 4 \pi F_T^3 \ , \]
 where $g_{ETC}$ and $M_{ETC}$ are the coupling and the energy scale of the ETC dynamics. In order to generate the top mass, it is necessary $M_{ETC} \approx g_{ETC}$ 1 TeV. If one considers $g_{ETC} \lesssim 4 \pi$, the cutoff scale for the top mass generation can be easily larger than 2-3 TeV. 
 }. From (\ref{YukawaU}) we get the following effective couplings between $a$ and the third quark generation 
 \begin{equation}
    \mathcal{L}_{\rm ferm} \;\supset\; -\frac{m_t}{F_T}\,i\,a\,\overline{t}\,\gamma_5\, t\,-\,\frac{m_b}{F_T}\,i\,a\,\overline{b}\,\gamma_5\, b\,\label{etaferm}\,,
 \end{equation}
which for $F_T=v$ predicts the effective coupling to the top $y_t=1$  in the effective Lagrangian (\ref{eq:L-eff}). The  Yukawa-like terms in \eqref{YukawaU}, at the underlying level, originate from effective four-fermion type interactions involving two techniquarks  and two SM fermions.  This scenario belongs to the class of models discussed in subsection~\ref{sec:GF}, where the diphoton
resonance is wide and it is produced via top-mediated gluon fusion. Furthermore we can neglect the extra top-mediated contributions to the effective coefficients $c_{VV}$ to the EW gauge bosons, since  they are subdominant compared to the gauged WZW terms given in \eqref{effc}.  

We report in the right plot of Fig.~\ref{fig:1}, the ${\rm BR}(a\to\gamma\gamma)$ versus $m_a$ in the case
the pseudoscalar resonance $a$ is coupled to the SM third generation according to (\ref{etaferm}), for $d(R)=6$ and different values of the techniquark hypercharge parameter $y$. The resulting total width
for $m_a=750$ GeV is $\Gamma_{\rm tot}(a)\approx$ 40 GeV, which is in remarkable agreement with the wide-width scenario reported by the ATLAS collaboration \cite{ATLAS}. In this case, taking 
$3.6 \lesssim y \lesssim 4.1$, it is possible to reproduce the diphoton excess\footnote{ Notice that in the case the pseudoscalar resonance is also coupled to a heavy vector-like quark, the gluon fusion production cross section can be enhanced reproducing the ATLAS excess at 750 GeV for even smaller values of $y$ \cite{Molinaro:2015cwg}.}. For values of the hypercharge parameter $y\approx 4 $ the diphoton branching ratio is of the order of 1\% at $m_a=750$ GeV. This implies that our scenario
is consistent with the ATLAS best-fit of the diphoton invariant mass, as discussed in subsection~\ref{sec:GF}
and shown explicitly in the left plot of Fig.~\ref{fig:ggF}. Similarly, the right plot of Fig.~\ref{fig:ggF} can be directly 
used to estimate the LHC-13 reach on the ($m_a$, $y$) parameter space. In particular, we can see that with a luminosity of 100  fb$^{-1}$ it is possible to test scenarios with ${\rm BR}(a\to\gamma\gamma)\lesssim 10\%$
($y\lesssim 4$) up to $m_a\approx 2$ TeV.\\
Finally, we may consider the case in which the heavy charged leptons have masses $m_L$ above $m_a/2$ generated by four-fermion interactions as for the other fermions\footnote{The current direct limit on the mass of heavy charged leptons is about $574$~GeV \cite{Chatrchyan:2013oca} for stable singly-charged particles.}. In this case, the new leptons couple to the pseudoscalar resonance as $-\frac{m_L}{F_T}i a \, \bar{L}\gamma_5 L$ and give a relevant contribution to the $a$ decay branching ratio to photons, lowering significantly the $y$ value needed to explain the diphoton excess. We find that for $d(R)=6$ the diphoton excess is reproduced for $y \approx 1.4$ for $m_L \gtrsim m_a$.

 \section{Conclusions}
\label{summary} 
The diphoton channel is  becoming increasingly attractive in spotting or constraining new physics at the LHC. Further inspired by the tantalizing excess at 750 GeV we  analyzed the reach of this channel first via an effective operator approach and then for minimal models of dynamical electroweak symmetry breaking. 

From our general effective analysis we deduced that it is possible to  directly photo-produce the 750 GeV resonance   without the need of any colored state and that in this case the resonance is narrow.  Beyond the 750 GeV excess, we have demonstrated that LHC-13 will constrain new diphoton resonances up to a wide energy range reaching circa 5 TeV with 100 fb$^{-1}$, as shown in Fig.~\ref{fig:aaF}. We also noted that a broad width for the 750 GeV resonance can be minimally achieved via a perturbative Yukawa-like coupling to the Standard Model top. In this case the resonance is produced via top-mediated gluon fusion and we showed in Fig.~\ref{fig:ggF} that the LHC reach is around 2 TeV for 100 fb$^{-1}$. 
 
In the second part of the paper we introduced minimal models of composite dynamics. Here the diphoton channel stems from  the topological sector of the theory and can account for the experimental excess. The relevant composite state is the pseudoscalar associated with the axial anomaly of the new underlying dynamics.  We independently analyzed the photon and the top-mediated gluon fusion production mechanisms and in both cases we determined the corresponding LHC reach. In the case of photon fusion we found that with 100 fb$^{-1}$ LHC-13 can probe values of the techniquark hypercharge parameter, defined in \eqref{HYtechniquarks}, as small as $y\approx 0.8$, as shown in Fig.~\ref{fig:aaF-EtaPrime}. For the top-mediated gluon fusion one can adopt the constraints found in the effective approach analysis, by using the matching relations in \eqref{effc}. In this case we obtained that with a luminosity of 100  fb$^{-1}$ LHC-13 can test values of $y \gtrsim 2$.  
Intriguingly, we noted that if  the Standard Model top mass is generated via  four-fermion operators, we can naturally explain the wide-width resonance reported by ATLAS.  

Our analysis shows that topological sectors from models of dynamical electroweak symmetry breaking leave imprints that can soon be tested by the LHC experiments.

\acknowledgments
 The CP$^3$-Origins center is partially funded by the Danish National Research Foundation, grant number DNRF90. 
 \appendix

\section{Wess-Zumino-Witten Action}
\label{}

The complete gauge invariant Wess-Zumino-Witten Lagrangian $\mathcal{L}_{\rm WZW}$, arising from the anomalous divergence of axial-vector currents in the underlying theory, can be expressed 
in terms of  the Maurer-Cartan one-forms
\begin{eqnarray}
\alpha & = & \left(\partial_{\mu}\mathcal{U}\right)\mathcal{U}^{-1} dx^{\mu}\equiv
\left(d\mathcal{U}\right)\mathcal{U}^{-1},~~~~
 \beta  \;=\;  
 \mathcal{U}^{-1}\alpha U~~~~~   \label{MC}
\end{eqnarray}
and additional ``left" and ``right" one-forms, $A_L=A^\mu_L dx_\mu$ and $A_R=A_R^\mu dx_\mu$, respectively, with $A^\mu_{L/R}$ defined in  (\ref{AR}).
We have
\begin{eqnarray}
 \int_{M^{4}}\mathcal{L}_{WZW} &=& \Gamma _{WZ}\left[ \mathcal{U}\right]
\,+\,5C\Bigg[ i\,\int_{M^{4}}{\rm Tr}\left[ A_{L}\alpha ^{3}+A_{R} \beta ^{3}\right]\,
-\,\int_{M^{4}}{\rm Tr}\left[ (dA_{L}A_{L}+A_{L}dA_{L})\alpha
+(dA_{R}A_{R}+A_{R}dA_{R})\beta \right]  \nonumber \\
&&+\,\int_{M^{4}}{\rm Tr}\left[
dA_{L}d\mathcal{U}A_{R}\mathcal{U}^{-1}-dA_{R}d\mathcal{U}^{-1}A_{L}\mathcal{U} \right] \,+\,\int_{M^{4}}{\rm Tr}\left[ A_{R}\mathcal{U}^{-1}A_{L}\mathcal{U}\beta
^{2}-A_{L}\mathcal{U}A_{R}\mathcal{U}^{-1}\alpha ^{2}\right]  \nonumber \\
&&+\,\frac{1}{2}\,\int_{M^{4}}{\rm Tr}\left[ (A_{L}\alpha
)^{2}-(A_{R}\beta )^{2}\right] \,+\,i\,\int_{M^{4}}{\rm Tr}\left[
A_{L}^{3}\alpha +A_{R}^{3}\beta \right] \nonumber\\
&&+\,i\,\int_{M^{4}}{\rm Tr}\left[
(dA_{R}A_{R}+A_{R}dA_{R})\mathcal{U}^{-1}A_{L}\mathcal{U}\,-\,(dA_{L}A_{L}+A_{L}dA_{L})\mathcal{U}A_{R}\mathcal{U}^{-1} \right]  \nonumber\\
&&+\,i\,\int_{M^{4}}{\rm Tr}\left[ A_{L}\mathcal{U}A_{R}\mathcal{U}^{-1}A_{L}\alpha
+A_{R}\mathcal{U}^{-1}A_{L}\mathcal{U}A_{R}\beta \right] \nonumber\\
&&+\,\int_{M^{4}}{\rm Tr}\left[
A_{R}^{3}\mathcal{U}^{-1}A_{L}\mathcal{U}-A_{L}^{3}\mathcal{U}A_{R}\mathcal{U}^{-1}+\frac{1}{2}
(\mathcal{U}A_{R}\mathcal{U}^{-1}A_{L})^{2}\right] \,-\, r\,\int_{M^{4}}{\rm Tr}\left[ F_{L}\mathcal{U}F_{R}\mathcal{U}^{-1}\right] \   \label{GWZW}
\end{eqnarray}
where $C=-i d(R)/(240\,\pi^2)$ and $r$ is a free parameter which is not determined by the gauge anomaly.
Here  $F_{L}$ and $F_{R}$ are  two-forms defined as
$F_{L}=dA_{L}-iA_{L}^{2}$ and $F_{R}=dA_{R}-iA_{R}^{2}$. 
The Wess-Zumino effective action is $\Gamma_{WZ}\left[\mathcal{U}\right]=C\, \int_{M^5} {\rm Tr} \left[\alpha^5\right]$, where $M^5$ is a five-dimensional manifold whose boundary $M^4$ denotes the Minkowski space.

\end{document}